# Tuning metal/superconductor to insulator/superconductor coupling via control of proximity enhancement between NbSe$_2$ monolayers

Olivio Chiatti[1], Klara Mihov[1,a], Theodor U. Griffin[1], Corinna Grosse[1,b], Matti B. Alemayehu[2], Kyle Hite[2], Danielle Hamann[2], Anna Mogilatenko[1,3], David C. Johnson[2] and Saskia F. Fischer[1,*]

The interplay between charge transfer and electronic disorder in transition-metal dichalcogenide multilayers gives rise to superconductive coupling driven by proximity enhancement, tunneling and superconducting fluctuations, of a yet unwieldy variety. Artificial spacer layers introduced with atomic precision change the density of states by charge transfer. Here, we tune the superconductive coupling between NbSe$_2$ monolayers from proximity-enhanced to tunneling-dominated. We correlate normal and superconducting properties in $[(SnSe)_{1+\delta})]_m[NbSe_2]_1$ tailored multilayers with varying SnSe layer thickness ($m = 1 - 15$). From high-field magnetotransport the critical fields yield Ginzburg-Landau coherence lengths with an increase of $140\%$ cross-plane ($m = 1 - 9$), trending towards two-dimensional superconductivity for $m > 9$. We show cross-over between three regimes: metallic with proximity-enhanced coupling ($m = 1 - 4$), disordered-metallic with intermediate coupling ($m = 5 - 9$) and insulating with Josephson tunneling ($m > 9$). Our results demonstrate that stacking metal mono- and dichalcogenides allows to convert a metal/superconductor into an insulator/superconductor system, prospecting the control of two-dimensional superconductivity in embedded layers.

[1] Novel Materials Group, Institut für Physik, Humboldt-Universität zu Berlin, Newtonstr. 15, 12489 Berlin, Germany
[2] Department of Chemistry and Materials Science Institute, University of Oregon, Eugene OR 97403, USA
[3] Ferdinand-Braun-Institut, Leibniz-Institut für Höchstfrequenztechnik, 12489 Berlin, Germany

[a] Present affiliation: Fraunhofer-Institut für Nachrichtentechnick, Heinrich-Hertz-Institut, HHI, Einsteinufer 37, 10587 Berlin, Germany
[b] Present affiliation: Berliner Nanotest und Design GmbH, Volmerstr. 9B 37, 12489 Berlin, Germany

# Introduction

Layered transition-metal dichalcogenides (TMDCs) have been popular materials to investigate as ultrathin layers, because they exhibit a wide range of electric transport properties, from insulating over metallic to superconducting [1-2]. Hence, they are well suited for the investigation of two-dimensional phenomena and for the development of new technologies [3]. The ability to stack single layers of different compounds and the discovery of topological materials have further spurred the study of new properties arising from proximity effects between adjacent layers [4-6]. This opens questions on the role played by changes in structure and bandstructure of spacer layers on the coupling between two-dimensional superconducting layers.

The two-dimensional structure of $NbSe_2$ lead researchers to probe for two-dimensional superconductivity by either cleaving crystals [7] or via intercalation of organic molecules [8]. The superconducting critical temperature decreased with either decreasing thickness or increasing separation of the $NbSe_2$ layers due to intercalation. More recently, two-dimensional superconductivity has been reported for $NbS_2$ layers in a crystalline $Ba_3NbS_5$ superlattice crystal [9], but the reports on superconductivity in $NbSe_2$ layers as a function of thickness or in heterostructures [10-16] do not paint a clear picture and leave open questions. Superconductivity with transition temperatures lower than the bulk have been observed in exfoliated and capped $NbSe_2$ monolayers (ML) [17-18] and in macroscopic encapsulated MLs grown by molecular beam epitaxy [19]. A systematic study for $NbSe_2$, where the inter-layer coupling can be changed continuously, is missing.

Generally, superconducting layers can be stacked with layers of either insulators, semiconductors or metals. The coupling between superconducting layers varies, depending on the properties and thickness of the other materials. When the spacer layers are insulating, the coupling between superconducting layers is due to tunneling [20]. For metallic spacer layers, the proximity effect needs to be considered [21-22]. A natural limitation of epitaxial structures and crystalline structures is a varying amount of strain induced by lattice mismatch as layer thicknesses are varied [23].

*Ferecrystals* [24] are novel artificially layered material systems, in which the individual layers are stacked by Van der Waals bonding with atomic precision and strain in growth direction is prevented by random orientation from layer to layer (turbostratic disorder). This allows to design and realize arbitrary stacking sequences [24].

In this work we investigate $[(SnSe)_{1+\delta})]_m[NbSe_2]_1$ ferecrystals, where $m = 1$ to 15, with SnSe-spacer layer thickness ranging from 0.583 nm to 8.72 nm. Due to the strain-free stacking of the NbSe₂ and SnSe layers, the in-plane lattice parameters of the NbSe₂ monolayers are constant as the SnSe layer thickness is increased [25-26]. Instead, the structure of the SnSe layers varies leading to charge transfer from SnSe to NbSe₂ [27]. The controlled insertion of SnSe layers in the ferecrystal films allows us to tune the superconductive coupling from strong to weak using the inter-layer spacing of the NbSe₂ monolayers, and to explore the onset of two-dimensional superconductivity for embedded monolayers.

The structural and electrical properties of these $[(SnSe)_{1+\delta})]_m[NbSe_2]_n$ ferecrystals samples are detailed in [28-32]. The parameter $\delta$ indicates the difference of the in-plane lattice parameter $a$ between the SnSe and NbSe₂ layers and its values range from 0.157 ($m = 1$) to 0.136 ($m = 9$) [28]. Structural details and electrical characterization above temperatures of 15 K of the compounds with $1 \leq m \leq 10$ and $n = 1$ were presented in [28]. A superconducting transition at low temperatures was reported for $1 \leq m \leq 6$ and $n = 1$ in [32], along with a detailed structural analysis, and [35] indicated that the coupling between NbSe₂ with increasing $m$ could lead to two-dimensional superconductivity, however that study was limited to $m \leq 9$, due to the availability of samples and cryogenic setups. Here, we include samples with $m \geq 9$ and extend the temperature range to below 100 mK.

This allows us to identify three transport regimes with varying inter-layer coupling, ranging from "good" metals to "dirty" metals to "bad" insulators, using the Ioffe-Regel criterion, and discuss them in term of charge transfer and disorder. The superconducting transition temperature decreases monotonically from 1.86 K to 0.25 K with increasing spacer layer thickness from 0.583 nm to 5.23 nm, consistent with a change from proximity-enhanced coupling to Josephson coupling.

## Methods

**Device fabrication.** The synthesis of the ferecrystals allows their structure to be tailored with atomic layer precision [24-29]. The technique uses a physical vapor deposition process, followed by annealing step, and is described in detail elsewhere [30-31]. Fig. **1**(a) gives a sketch to illustrate the layer sequence for the $[(SnSe)_{1+\delta})]_m[NbSe_2]_1$ ferecrystal with $m = 1$: between each $[NbSe_2]_1$ *mono*layer there are $m$ SnSe layers, separated by Van der Waals (VdW) gaps. For the electrical transport measurements, ferecrystals were deposited through shadow masks in a cross, cloverleaf or Hall-bar shape on insulating substrates. These geometries allow to perform four-point Van der Pauw (VdP) and Hall measurements. The substrates were glued to chip carrier with silver paste and thin gold wires were attached to the contact pads with small Indium pieces.

**Structural investigations.** Fig. **1**(b) shows specular X-ray diffraction (XRD) patterns of the $[(SnSe)_{1+\delta})]_m[NbSe_2]_1$ ferecrystals with $m = 6, 9, 12, 15$, which demonstrate the precision of the layer sequences [31-32]. This precision is confirmed by high-angle annular dark-field scanning transmission electron microscopy (HAADF-STEM). Fig. **1**(c) shows a typical HAADF-STEM image of the ferecrystal with $m = 6$. This image also shows examples of the turbostratic disorder (white arrows), where the crystal orientation of the grains changes within a layer and between adjacent layers. The thickness of the $[NbSe_2]_1$ monolayers is $t_{NbSe_2} = 0.6675$ nm and the thickness of the $[(SnSe)_{1+\delta}]_m$ spacer layers is $t_{SnSe} = m \cdot 0.58125$ nm.

**Electric transport measurements.** Low temperatures down to 50 mK were achieved in an *Oxford Instruments Triton* dilution refrigerator, with a superconducting magnet for magnetic fields up to 12 T. The chip carriers can be mounted in the sample holder either perpendicular or parallel to the magnetic field, before inserting the sample holder into the cryostat for cooldown. The orientation cannot be changed after insertion. The bath temperature $T_{\text{bath}}$ is determined below 2 K with a calibrated Ruthenium-Oxide resistance thermometer, located between the sample holder and the mixing chamber. The superconducting coil generates magnetic fields up to 12 T. In general, the field was swept in a quasi-static way, keeping the field constant for several minutes before collecting data, to allow for the sample resistance to settle at a constant value for the given field. This time-dependence of the magnetoresistance yields a hysteresis, which is particularly pronounced in the parallel field configuration for the ferecrystal with $m = 9$. Resistance measurements were performed in two ways. First, the current-voltage characteristics for different contact configurations were measured with a *Keithley* 6221 current source and a 2182A nanovoltmeter, using the VdP method to determine the sheet resistance [33-34]. Second, four-point resistance measurements were performed with lock-in amplifiers (*Signal Recovery* 7265 or *Stanford Research Systems* SR830), using currents between 1 nA and 50 nA at frequencies from 12 Hz up to 433 Hz.

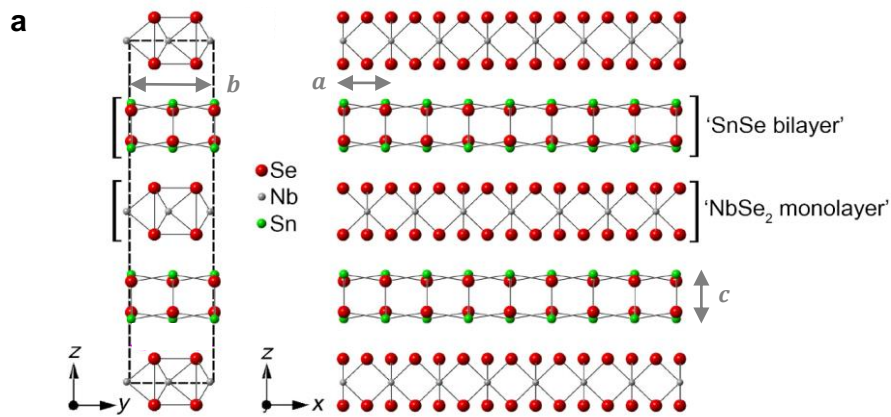

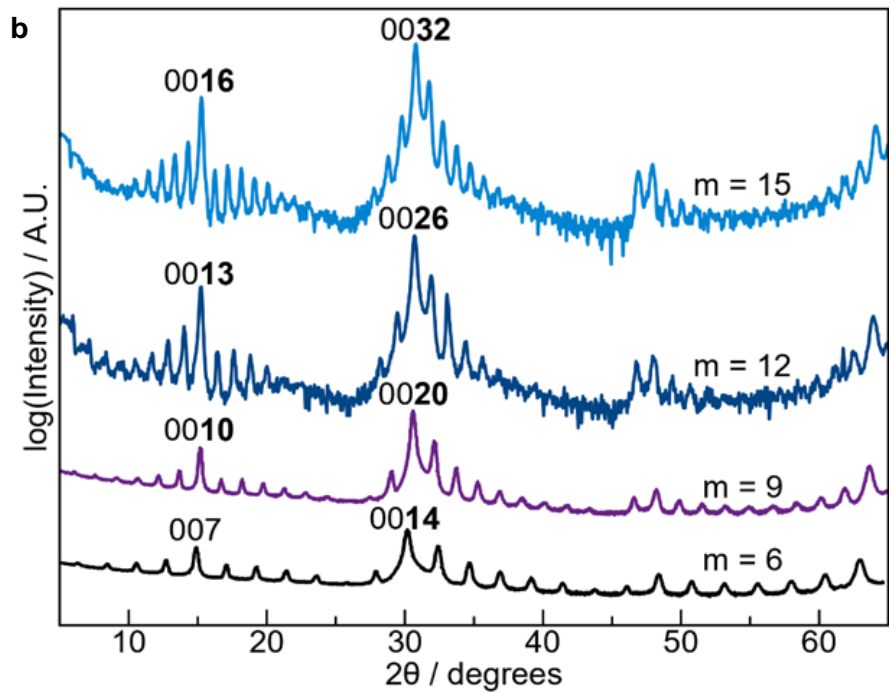

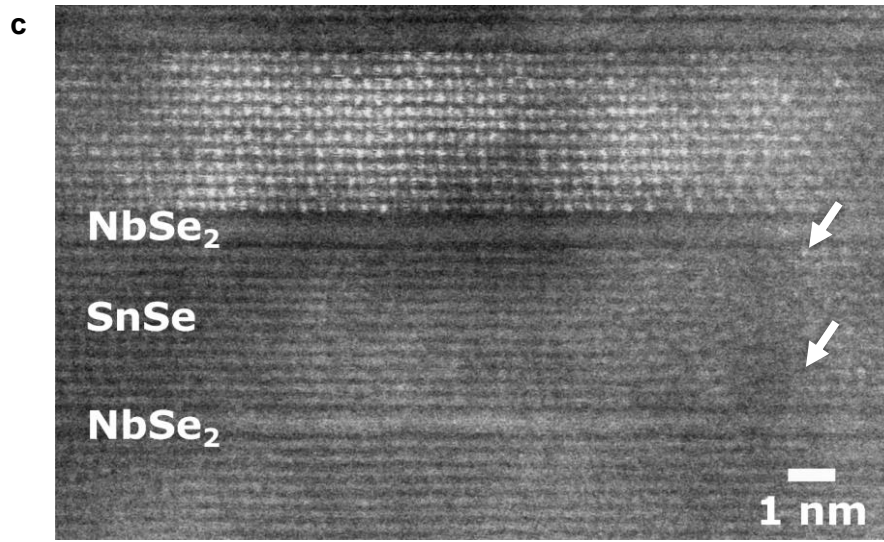

**Fig. 1**

**a** Sketch of the layer sequence of the $[(SnSe)_{1+\delta}]_m(NbSe_2)_1$ ferecrystal with $m = 1$. The layers are projected onto the $yz$-plane (left) and onto the $xz$-plane (right). $a$, $b$, and $c$ denote the lattice parameters within the layers.

**b** Specular X-ray diffraction patterns of the $[(SnSe)_{1+\delta}]_m(NbSe_2)_1$ ferecrystals with $m = 6, 9, 12$ and $15$. The patterns are vertically offset from one another for clarity and two reflections in each pattern are indexed with the corresponding Miller indices.

**c** High-resolution HAADF-STEM image of the $[(SnSe)_{1+\delta}]_m(NbSe_2)_1$ ferecrystal with $m = 6$. It shows that he deposition technique allows for monolayer precision of the layers. The arrows indicate regions within a layer where the crystal orientation is different, illustrating the turbostratic disorder.

## Results

**Temperature dependence of sheet resistance.** Fig. 2 shows the results of the temperature-dependent measurements of the resistance for the $[(\text{SnSe})_{1+\delta}]_m[\text{NbSe}_2]_1$ ferecrystals with $m = 1, 3, 6, 9, 12, 15$. The sheet resistance per superconducting layer $R_s^* = R_s \cdot M_{\text{NbSe}_2}$ shows a superconducting transition with $T_c > 0$ for $m = 1, 3, 6$ and $9$. The values for $T_c$ are determined at $R_s = 50\% \cdot R_{s,N}$, where $R_{s,N}$ is the sheet resistance in normal-state above the transition, in order to account for a broadening of the transition due to thermal fluctuations. The sheet resistance for $m = 12$ and $15$ shows a qualitatively different behavior: at temperatures below a crossover temperature $T_{co}$, the resistance becomes non-monotonic, with a local minimum at $T_{\text{bath}} = (0.12 \pm 0.1)$ K and $(0.20 \pm 0.02)$ K for $m = 12$ and $15$, respectively (see Fig. **3**(a)). This is reminiscent of the quasi-

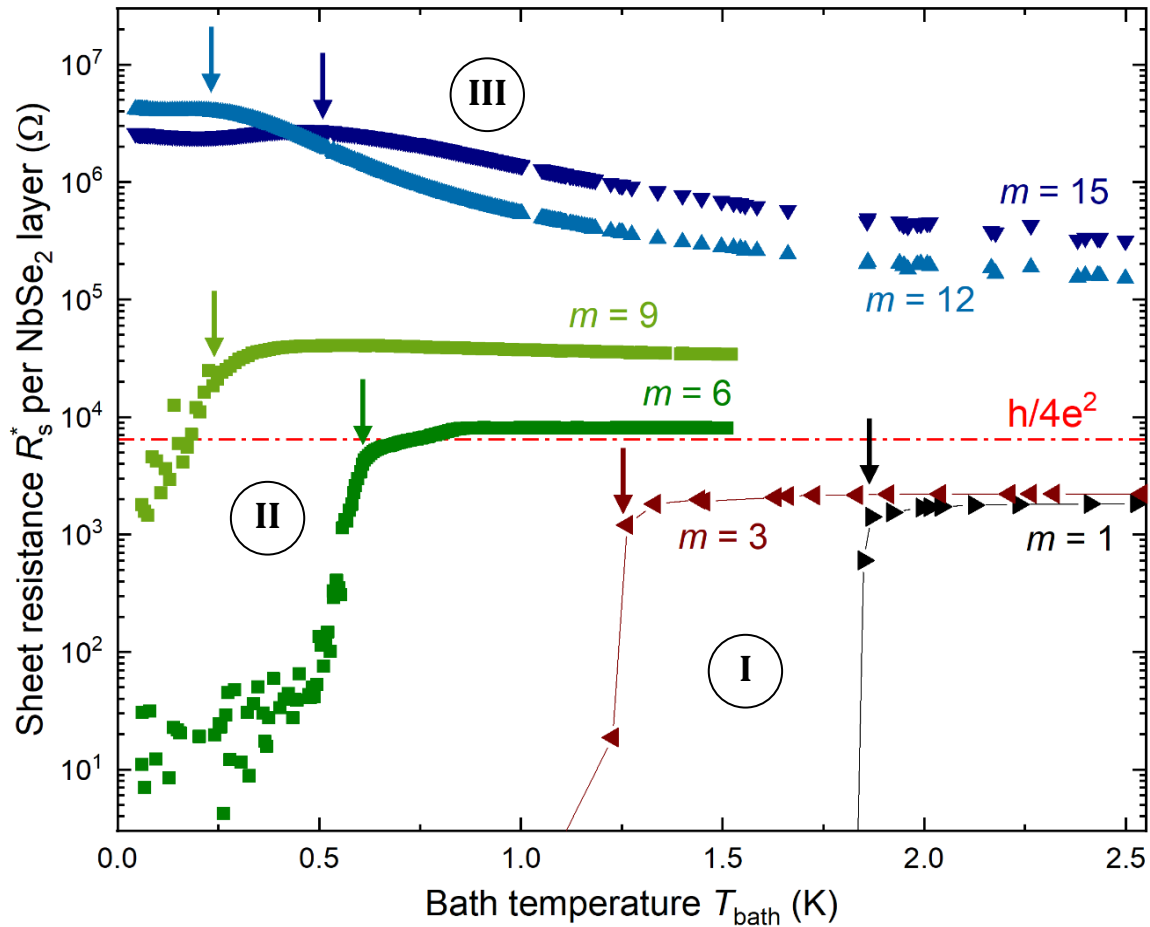

**Fig. 2** Low-temperature sheet resistance $R_s^*$ per $(\text{NbSe}_2)_1$ monolayer of the $[(\text{SnSe})_{1+\delta}]_m(\text{NbSe}_2)_1$ ferecrystals with $m = 1, 3, 6, 9, 12$ and $15$. The arrows indicate either the transition temperature $T_c$ ($m = 1, 3, 6, 9$) or the crossover temperature $T_{co}$ ($m = 12, 15$). The lines for $m = 1, 3$ are guides to the eye. The dash-dotted line corresponds to a resistance of $h/4e^2 \approx 6.5$ kΩ. The roman numerals denote different physical origins of the transport behavior, as discussed in the section 'Discussion', second paragraph. Data for $m = 1, 3, 6$ are from our previous work [31].

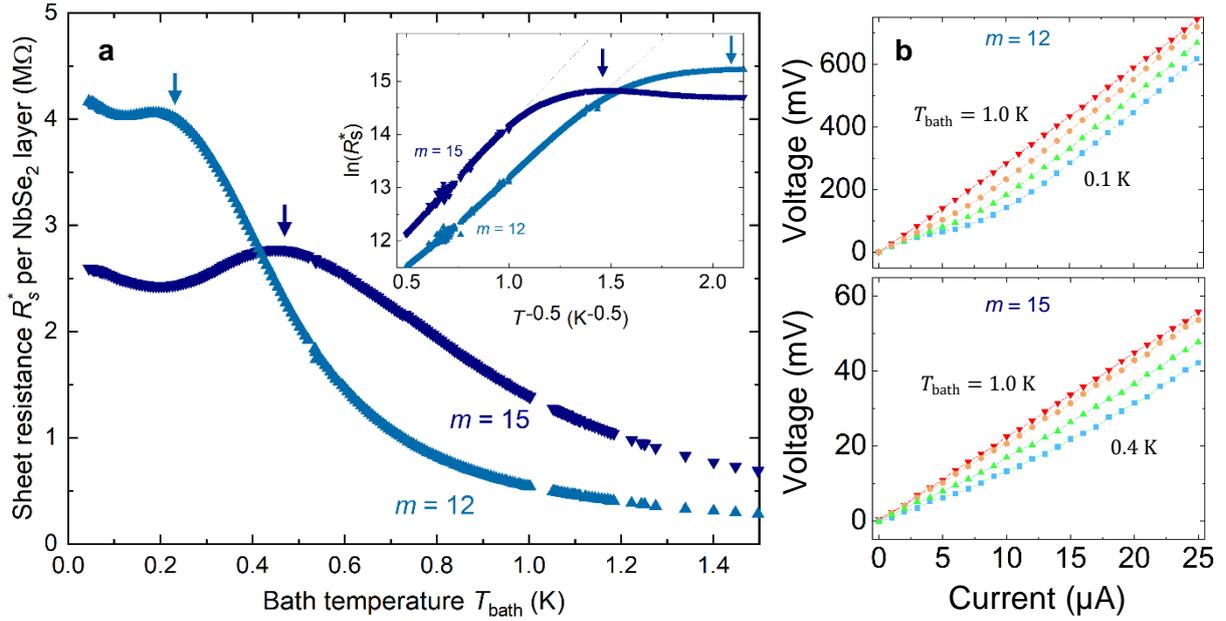

**Fig. 3 a** Sheet resistance $R_s^*$ per $[NbSe_2]_1$ monolayer of the ferecrystals with $m = 12$ and 15. The arrows indicate the crossover temperatures $T_{co}$. Inset: Plot of $\ln(R_s^*)$ as a function of $T^{-1/2}$. A straight line is consistent with transport by variable-range hopping (VRH): $R_s^*(T) = R_{s,0}^* \exp\left[\left(\frac{T_{VRH}}{T}\right)^{1/2}\right]$. For $m = 12$ VRH is found for $T_{bath} > 1.0$ K ($T_{VRH} = 11.6$ K); for $m = 15$, $T_{bath} > 0.5$ K ($T_{VRH} = 16.7$ K). **b** Exemplary $IV$-curves of the ferecrystals with $m = 12$ and 15, for different temperatures. The non-linearity at lower temperatures indicates the presence of a supercurrent [39].

reentrant behavior of homogeneously disordered ultra-thin superconducting films [36]. Above $T_{co}$ the temperature dependence is consistent with variable-range hopping [37] (see inset of Fig. **3**(a)). The temperature dependence and the current-voltage characteristics (see Fig. **3**(b)) are both consistent with the onset of a "local" superconductivity in the NbSe₂ layers below $T_{co}$ [36, 38].

**Magnetic-field dependence of sheet resistance.** The upper critical field $B_{c2}$ depends on the temperature through the temperature-dependent Ginzburg-Landau (GL) coherence length $\xi(T)$ [20, 40]. In highly anisotropic materials the GL-coherence length depends on the orientation in the material. In particular, due to the layered structure of the ferecrystals, the GL coherence length parallel to the layers $\xi_{ab}$ can be assumed to be isotropic within the plane parallel to the layers. In a perpendicular magnetic field, the superconducting properties depend only on $\xi_{ab}$ [20, 40-42], and the upper critical field near the zero-field transition temperature $T_c$ can be written as follows:

$$B_{c2,\perp} = \frac{\Phi_0}{2\pi \xi_{ab}^2(T)} = \frac{\Phi_0}{2\pi \xi_{ab}^2(0)}\left(1 - \frac{T}{T_c}\right), \tag{3.1}$$

with the flux quantum $\Phi_0 = 2.07 \times 10^{-15}$ Tm² and the temperature-dependent GL-coherence length $\xi_{ab}(T) = \xi_{ab}(0)\left(1 - \frac{T}{T_c}\right)^{-\frac{1}{2}}$.

In a parallel magnetic field, the temperature dependence of the upper critical field near $T_c$ can be described by the Lawrence-Doniach model in the anisotropic GL-limit for fully coupled layers (3D) [20]:

$$B_{c2,\parallel} = \frac{\Phi_0}{2\pi\xi_{ab}(T)\xi_c(T)} = \frac{\Phi_0}{2\pi\xi_{ab}(0)\xi_c(0)}\left(1 - \frac{T}{T_c}\right). \qquad (3.2)$$

Both equations (3.1) and (3.2) yield a linear dependence of the parallel and perpendicular critical fields on temperatures near the transition temperature. This dependence is observed in Fig. **4**(a), (c) and (d), where the parallel and perpendicular critical fields are shown in dependence on the temperature. The linear dependence close to the critical temperature can be fitted to calculate both coherence lengths of the anisotropic ferecrystals. The resulting coherence lengths are shown in in comparison to the repeat unit thickness $s$. The distance $s$ was determined by X-ray diffraction [28] and confirmed by STEM and selected-area electron diffraction [31].

**Table 1** Superconducting transition temperature $T_c$, number of NbSe₂ layers per film $M_{\text{NbSe}_2}$, room-temperature sheet resistance $R_{s,300\,K}$, low-temperature ($T \geq T_c$) normal-state resistance $R_{s,N}$, and low-temperature normal-state Hall coefficient $R_{H,N}$ of the $[(\text{SnSe})_{1+\delta}]_m(\text{NbSe}_2)_1$ ferecrystals with $m = 1 - 15$. For $m = 12$ and 15 a transition temperature estimated from the disappearance of the non-linearity in the $IV$-curves is given in round brackets; for comparison, the cross-over temperature $T_{co}$ is given in square brackets.

| $m$ | Transition temperature $T_c$ (K) | Total number of NbSe₂ monolayers per film $M_{\text{NbSe}_2}$ | Room-temperature sheet resistance $R_{s,300\,K}$ (Ohm) | Low-temperature normal-state sheet resistance $R_{s,N}$ (Ohm) | Low-temperature normal-state Hall coefficient $R_{H,N}$ (Ohm m/T) |
|---|---|---|---|---|---|
| 1 | 1.86 ± 0.05 | 37 | 82.47 ± 0.05 | 50.54 ± 0.04 | (1.444 ± 0.002)×10⁻⁹ |
| 3 | 1.26 ± 0.03 | 22 | 146.7 ± 0.2 | 100.5 ± 0.2 | (3.52 ± 0.02)×10⁻⁹ |
| 5 | 0.81 ± 0.05 | 11 | 471.8 ± 0.5 | 484.0 ± 0.5 | (8.11 ± 0.05)×10⁻⁹ |
| 6 | 0.61 ± 0.08 | 9 | 754.7 ± 0.7 | 950.0 ± 0.8 | (11.48 ± 0.03)×10⁻⁹ |
| 9 | 0.25 ± 0.02 | 7 | 1455 ± 1 | 5800 ± 0.4 | (53 ± 22)×10⁻⁹ |
| 12 | (0.45 ± 0.09) [0.23 ± 0.05] | 9 | 2340 ± 3 | 450500 ± 500 | - |
| 15 | (0.51 ± 0.09) [0.47 ± 0.05] | 9 | 2821 ± 3 | 308000 ± 400 | - |

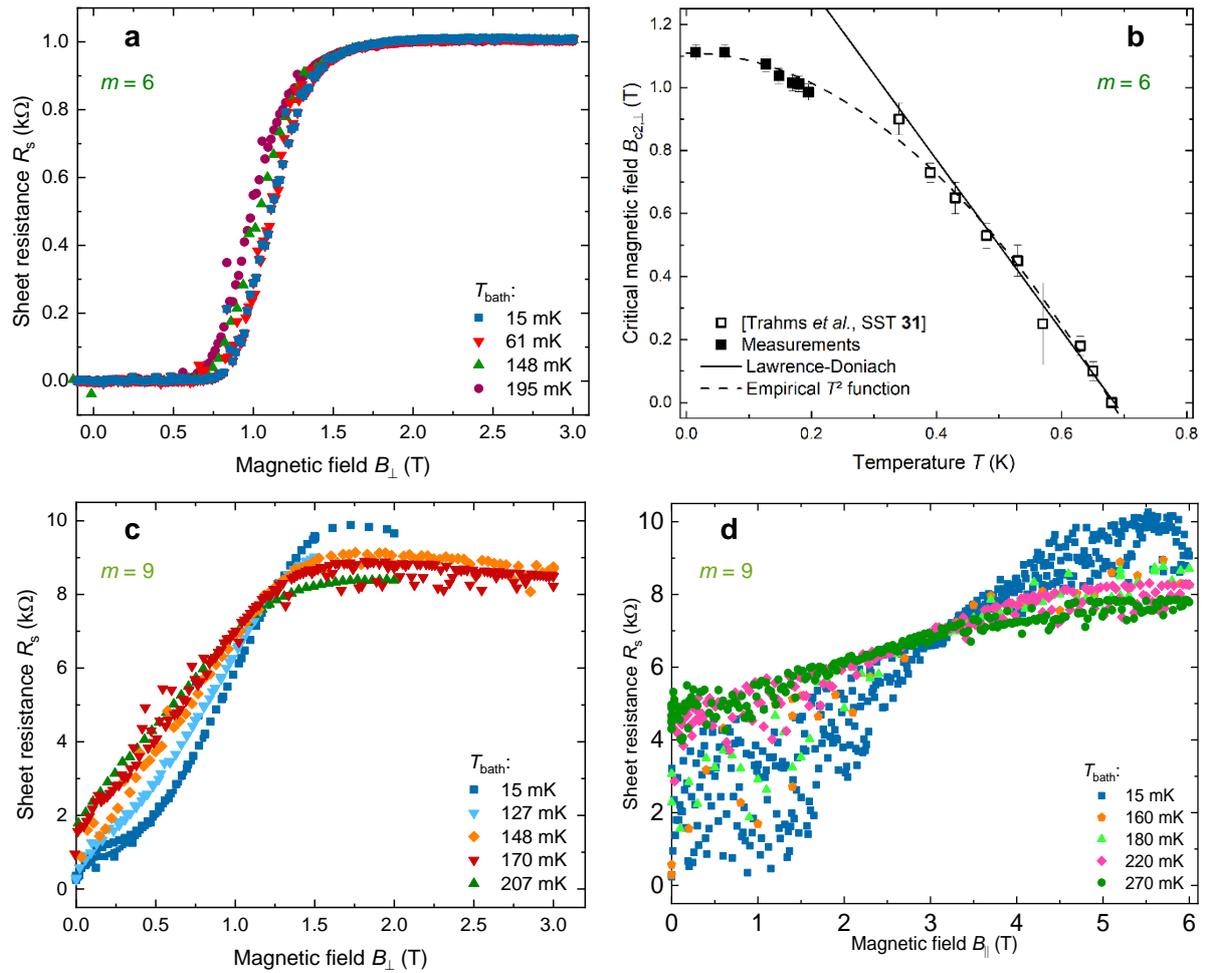

**Fig. 4 a** Resistance as a function of perpendicular magnetic field of the $[(SnSe)_{1+\delta})]_m[NbSe_2]_1$ ferecrystal with $m = 6$. **b** Upper critical (perpendicular) field as a function of temperature, obtained with the 50% · $R_{s,N}$ criterion, showing data from measurements in this and previous works [35]. The solid line is a fit with the Lawrence-Doniach model (Eq. 3.1); the dashed line is a guide-to-the-eye based on an empirical $T^2$-function. **c** Sheet resistance of the $[(SnSe)_{1+\delta})]_m[NbSe_2]_1$ ferecrystal with $m = 9$ as a function of a magnetic field perpendicular to the film, for varying bath temperatures. (d) Sheet resistance for $m = 9$ with magnetic field parallel to the film: the strong fluctuations indicate the presence of vortices spanning both the superconducting and non-superconducting layers.

Fig. **4**(a) shows the sheet resistance of the ferecrystal with $m = 6$ as a function of the perpendicular magnetic field $B_\perp$. A transition from the superconducting to the normal state is observed for bath temperatures $T_\text{bath} < T_c$. The upper critical field $B_{c2}$ is obtained using the 50% · $R_{s,N}$ criterion. Fig. **4**(b) shows the upper critical fields $B_{c2,\perp}(T)$ of the ferecrystal with $m = 6$ in a perpendicular magnetic field $B_\perp$ from Fig. **4**(a). The present results are consistent with previous results on the same samples [35]: current and previous data fall onto the same empirical $T^2$ function. A fit with the Lawrence-

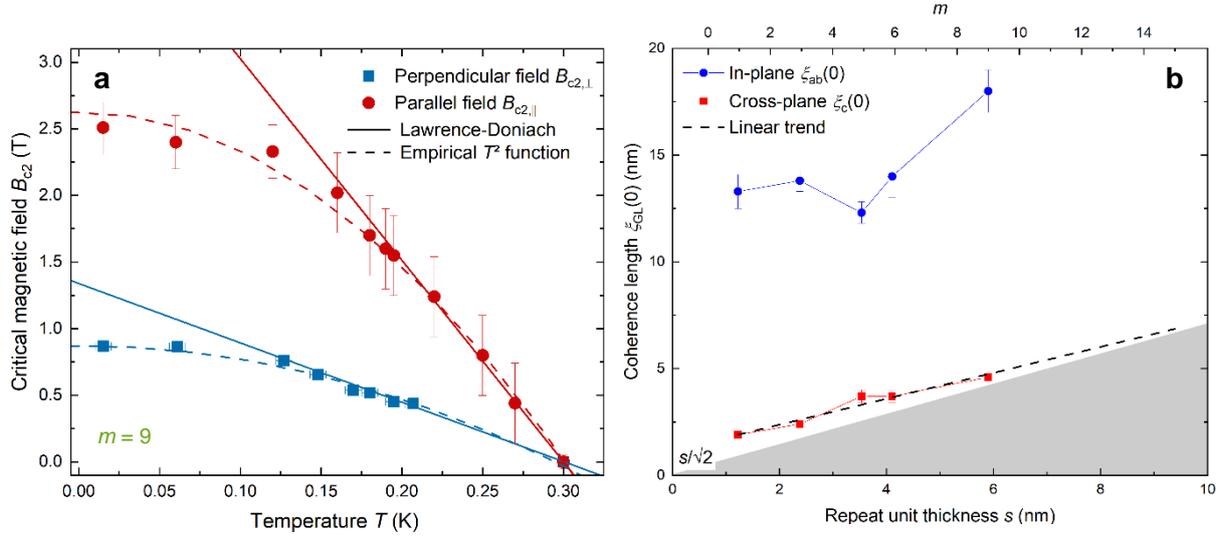

**Fig. 5 a** Upper critical fields of the $[(\text{SnSe})_{1+\delta}]_m[\text{NbSe}_2]_1$ ferecrystal with $m = 9$ as a function of temperature, obtained with the $50\% \cdot R_{s,N}$ criterion. The solid lines are fits with the Lawrence-Doniach model; the dashed lines are guides-to-the-eye based on an empirical $T^2$-function. **b** In-plane and cross-plane coherence lengths as a function of repeat unit thickness for the $[(\text{SnSe})_{1+\delta}]_m[\text{NbSe}_2]_1$ ferecrystals with $m = 1 - 9$. A linear extrapolation of the cross-plane coherence lengths yields $\xi_c(0) = \frac{s}{\sqrt{2}}$ for $s \approx 11.6$ nm, or $m = 18 - 19$.

Doniach model (Eq. 3.1) of $B_{c2,\perp}(T)$ for temperatures near $T_c$ yields the in-plane GL coherence length $\xi_{ab}(0)$.

The sheet resistance of the ferecrystal with $m = 9$ as a function of the perpendicular magnetic field $B_\perp$ is shown in Fig. **4**(c). A transition from the superconducting to the normal state is observed for bath temperatures $T_{\text{bath}} < T_c$. The upper critical field $B_{c2,\perp}$ is obtained using the $50\% \cdot R_{s,N}$ criterion. The sheet resistance in parallel magnetic field $B_{\parallel}$ in Fig. **4**(d) also shows a transition. However, the resistance appears to fluctuate strongly with varying field. This is not observed for $m \leq 6$ [35]. The amplitude of these fluctuations decreases with increasing temperatures. This behavior is similar to the universal conductance fluctuations in mesoscopic conductors: here the microscopic configuration of disorder has a strong effect on the measured resistance and can change with varying external fields. With increasing temperature, it is averaged out. In our measurements, the pancake vortices of the magnetic flux thread the layered structure and strongly affect the resistance of the few vortices are partly within the superconducting layers and partly superconducting layers. In the limit of magnetic field parallel to the layers, the supercurrents of the in the non-superconducting layers [20]. The parts in the proximity-affected SnSe layers are more affected by disorder.

The spatial configuration of the vortices changes with varying magnetic field and is "averaged out" with increasing temperature. It is possible to estimate the upper critical field $B_{c2,\parallel}$ using the $50\% \cdot R_{s,N}$ criterion.

Fig. **5**(a) shows the upper critical fields $B_{c2,\perp}(T)$ and $B_{c2,\parallel}(T)$ of the ferecystal with $m = 9$ from Fig. **4**(c) and (d). All data fall onto empirical $T^2$ functions. Fits with the Lawrence-Doniach model of $B_{c2,\perp}(T)$ (Eq. 3.1) and $B_{c2,\parallel}(T)$ (Eq. 3.2) for temperatures near $T_c$ yield the in-plane and cross-plane Ginzburg-Landau coherence lengths $\xi_{ab}(0)$ and $\xi_c(0)$. The results of the fits are summarized in Table 2 and compared to the repeat unit thickness $s$. Table 2 also includes results from previous works [31, 35].

The in-plane and cross-plane coherence lengths show linear trends as a function of increasing repeat unit thickness for the ferecrystals with $m = 1 - 9$ (see Fig. **5**(b)). The anisotropy $\xi_{ab}(0)/\xi_c(0)$ decreases non-monotonically with increasing $s$ and "saturates" at 3.9 for $m = 9$. Extrapolating the linear trend of $\xi_c(0)$, we find that $\xi_c(0) = \frac{s}{\sqrt{2}}$ for $s \approx 11.6$ nm, which corresponds to $m = 18 - 19$.

**Table 2** Repeat unit thickness $s$, in-plane and cross-plane coherence lengths, $\xi_{ab}(0)$ and $\xi_c(0)$, anisotropy and 2D-criterion ratio for the $[(\mathrm{SnSe})_{1+\delta}]_m[\mathrm{NbSe}_2]_1$ ferecrystals with $m = 1 - 9$ (includes results from previous works [31, 35]).

| $m$ | Repeat unit thickness $s$ (nm) | In-plane coherence length $\xi_{ab}(0)$ (nm) | Cross-plane coherence length $\xi_c(0)$ (nm) | Anisotropy $\xi_{ab}(0)/\xi_c(0)$ | Ratio $\xi_c(0)/\frac{s}{\sqrt{2}}$ |
|---|---|---|---|---|---|
| Single crystal [40-43] | 0.60 | 9.0 - 11.0 | 2.7 - 4.0 | 2.2 – 4.1 | ≈ 8 |
| 1 | 1.223 ± 0.001 | 13.3 ± 0.8 | 1.9 ± 0.1 | 7.0 ± 0.4 | 2.2 ± 0.1 |
| 3 | 2.378 ± 0.001 | 13.8 ± 0.5 | 2.4 ± 0.1 | 5.8 ± 0.2 | 1.4 ± 0.1 |
| 5 | 3.533 ± 0.001 | 12.3 ± 0.5 | 3.7 ± 0.5 | 3.3 ± 0.5 | 1.5 ± 0.2 |
| 6 | 4.106 ± 0.001 | 14 ± 1 | 3.7 ± 0.6 | 3.8 ± 0.6 | 1.3 ± 0.2 |
| 9 | 5.9 ± 0.2 | 18 ± 1 | 4.6 ± 0.1 | 3.9 ± 0.1 | 1.10 ± 0.05 |
| 12 | 7.6 ± 0.2 | - | - | - | - |
| 15 | 9.4 ± 0.2 | - | - | - | - |

## Discussion

**The crystal structure and the electronic bandstructure** of $[(SnSe)_{1+\delta})]_m[NbSe_2]_1$ ferecrystals established by detailed analysis via XRD and photoemission spectroscopy (PES) [26-28] indicate that the structure of the SnSe layers varies with increasing $m$, with an increase of the in-plane $a$ lattice parameter of 2.3% from $m = 1$ to 10. Bandstructure calculations of mono- and bilayer SnSe [47] have shown a reduction in the bandgap of approximately 10%. The structure of the NbSe$_2$ monolayers is not as strongly affected by $m$, with an increase of the $a$ lattice parameter of 0.3%. Photoelectron spectroscopy of $[(SnSe)_{1+\delta})]_m[NbSe_2]_2$ ferecrystals [27] showed that there is a charge transfer of electrons from SnSe to NbSe$_2$. A comparison of electric transport properties between $[(SnSe)_{1+\delta})]_m[NbSe_2]_1$ and $[(SnSe)_{1+\delta})]_m[NbSe_2]_2$ ferecrystals [48] concluded that the charge transfer scales with the number of NbSe$_2$ layers. Therefore, the structural change of the SnSe layers affects the charge transfer to the NbSe$_2$ layers through a change of the density of states (DOS) in the SnSe layers. Hence, in a first approximation the electric transport properties of $[(SnSe)_{1+\delta})]_m[NbSe_2]_1$ ferecrystals with varying $m$, and in particular the coupling between the superconducting NbSe$_2$ layers, are determined by the SnSe layers. Further evidence comes from magnetotransport measurements, as the in-plane coherence length changes by 13% from $m = 1$ to 9, while the cross-plane coherence length changes by 140%, indicating that the spacer layer affects primarily the inter-layer coupling.

The **temperature dependence of the sheet resistance** depends on the variation of the inter-layer coupling with spacer layer thickness. For the $[(SnSe)_{1+\delta})]_m[NbSe_2]_1$ ferecrystals with $m = 1 - 15$ this suggests qualitatively the identification of three regimes (I, II, III in Fig. **2**) for the transport properties as the number of SnSe layers $m$, or the repeat unit thickness $s$, is increased:

In **regime I** ($m \leq 3$, Fig. **2**) the temperature dependence of the sheet resistance is metallic-like. The variation of the electronic bandstructure of mono- and bilayer SnSe [47] yields a charge transfer between SnSe layers and NbSe$_2$ layers [28], which differs significantly between one, two and three SnSe layers. This also affects the room-temperature sheet resistance, the temperature coefficient of $R_s(T)$ and the superconducting transition temperature $T_c$: due to the charge transfer, the chemical potential in the SnSe layers lies in the valence band, making the spacer layer a degenerate semiconductor with a finite DOS. This makes a proximity effect with the NbSe$_2$ layers possible, which enhances the inter-NbSe$_2$-layer coupling. The cross-plane GL coherence length is larger than the inter-NbSe$_2$-layer spacing, consistent with an anisotropic 3D superconductor, and a macroscopic superconducting state is established below $T_c$.

In **regime II** ($4 \leq m \leq 9$, Fig. **2**) the temperature dependence of the sheet resistance is semi-conducting-like. The bandstructure of the SnSe layers is bulk-like [47] and the change of the charge transfer with increasing the number of SnSe layers is much smaller than in regime I [28]. The sheet resistance is one to two orders of magnitude higher compared to $m \leq 3$, with a nearly exponential dependence on the repeat unit thickness. This suggests that the enhancement of the inter-NbSe$_2$-layer coupling due to proximity effects is reduced, through the decrease of the DOS in the semiconducting SnSe layer. The onset of superconductivity occurs at lower temperatures due to the decreasing three-dimensional character of superconducting fluctuations with increasing separation of the NbSe$_2$ layers. This is consistent with the cross-plane coherence length being close to the repeat unit thickness.

In **regime III** ($9 \leq m$, Fig. **2**) the temperature dependence of the sheet resistance is more insulating-like with variable range hopping above a cross-cover temperature. The transition to a global superconducting state does not appear to occur at a finite temperature. In this regime the separation of the NbSe$_2$ layers is large enough to reduce the inter-layer coupling to Josephson tunneling, with the cross-plane coherence length smaller than the separation of the superconducting layers. The intra-layer scattering due to disorder is increased, because screening in the SnSe layers is reduced. Therefore, the granular structure of the NbSe$_2$ layers and the reduced inter-layer coupling result in variable-range hopping transport in the normal state and the onset of superconductivity is only locally observable, in the form of a quasi-reentrant effect in the temperature-dependent sheet resistance and non-linear $IV$-curves at low temperatures. The non-linearity observed in the $IV$-curves is similar to that of 2D Josephson-junctions arrays near the superconductor-insulator transition [39] and indicates the presence of a supercurrent, even without a macroscopic superconducting path through the array. We have not observed a clear signature of the BKT transition [40], as in zero magnetic field the current-voltage characteristics do not show the expected $V \propto I^3$ behavior and in magnetic fields the characteristic

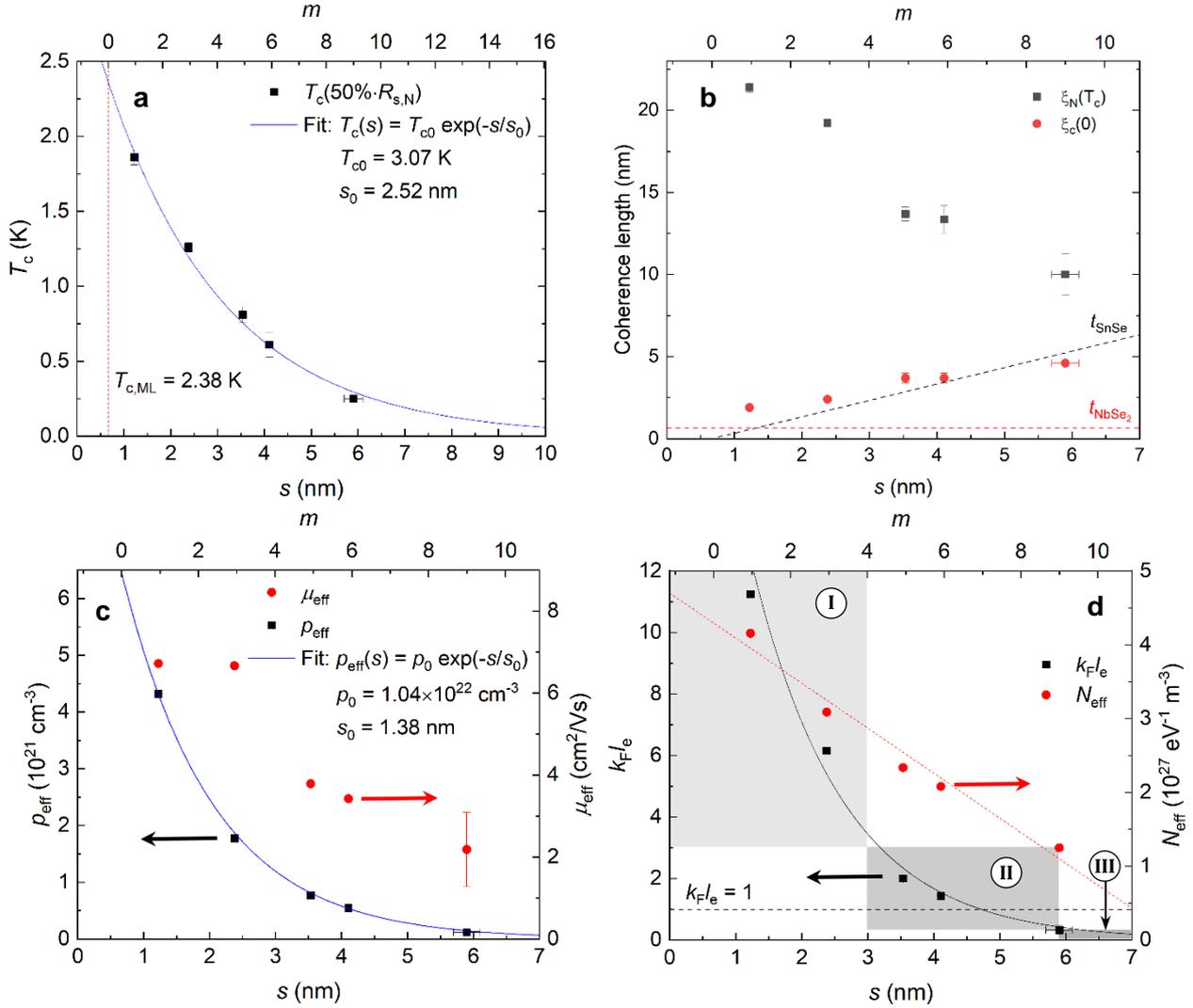

**Fig. 6 a** The superconducting transition temperature $T_c$ as a function of the repeat unit thickness $s$ for the $[(SnSe)_{1+\delta}]_m[NbSe_2]_1$ ferecrystals with $m = 1 - 9$, estimated with the $50\% \cdot R_{s,N}$ criterion (black squares). The solid blue line is a fit to an exponential function (see text and Eq. 4.3-4.5). The dashed red line is the extrapolated transition temperature for a NbSe$_2$ monolayer. **b** Cross-plane superconducting coherence length $\xi_c(0)$ (Table 2) for the NbSe$_2$ layer (red circles) and diffusive coherence length near $T_c$, $\xi_N(T_c)$ (Eq. 4.6) for the SnSe layer (black squares), as a function of repeat unit thickness $s$. The dashed lines are the thicknesses of the respective layers in a repeat unit. **c** Effective charge carrier density $p_{eff}$ (black squares) and effective Hall mobility $\mu_{eff}$ (red circles), calculated from the low-temperature resistivity and Hall-coefficient in the normal state, as a function of $s$. The solid blue line is a is a fit of $p_{eff}(s)$ to an exponential function (see text). **d** Product of Fermi wavelength $k_F$ and mean free path $l_e$ for the Ioffe-Regel criterion (black squares) and effective density of states $N_{eff}$ (red circles), as a function of $s$ (see text and Eq. 4.11). The solid black line is a guide to the eye for $k_F l_e$; the dashed black line indicates $k_F l_e = 1$. The short-dashed red line is a guide to the eye for $N_{eff}$. The roman numerals denote different physical origins of the transport behavior, discussed in the section 'Discussion', second paragraph.

fields for a cross-over from 3D- to 2D-melting of a flux-line lattice is larger than $B_{c2}$ for all samples. However, the grain boundaries within the layers provide pinning centers for vortices, so that an observation of flux-line flow is not expected.

In **superconductor-insulator** (SI) multilayers the transport between the superconducting layers is dominated by tunneling and the strength of the coupling depends exponentially on the inter-layer separation $s$, compared to the Ginzburg-Landau (GL) coherence length perpendicular to the layers (cross-plane) $\xi_c(0)$. The ratio $\xi_c(0)/s$ measures the strength of the inter-layer coupling, which determines whether the multilayer behaves as an anisotropic 3D superconductor or as a stack of 2D superconductors. This can lead to a dimensional crossover from 3D to quasi-2D superconductivity if the GL coherence lengths are varied. The criterion for 2D layers with weak Josephson inter-layer coupling is given by [20]

$$\xi_c(0) < \frac{s}{\sqrt{2}}. \qquad (4.1)$$

Table 2 shows that the ratio $\xi_c(0)/(s/\sqrt{2})$ decreases with increasing $m$, with a linear trend towards 1 for $s \approx 11.6$ nm, or $m \approx 18 - 19$. Based only on this criterion, for $m > 18$ it is expected that the superconducting layers are fully decoupled. Therefore, the ferecrystals with $m = 1$ to 15 should be considered as anisotropic 3D superconductors, with a trend towards 2D-superconductivity by decoupling due to increasing separation of the NbSe₂ layers. If the structure and bandstructure of the layers do not change with $s$, then the superconducting transition temperature is expected to vary exponentially with the inter-layer pairing interaction, which in turn depends exponentially on the inter-layer separation $s$ [20, 49].

In **superconductor-normal metal** (SN) multilayers the transport between superconducting layers is dominated by normal, metallic transport [21-22]. The non-superconducting layers act as weak links between the superconducting layers and the inter-layer separation $s$ can control the strength of the inter-layer coupling. Due to the proximity of the superconducting layers the order parameter in the normal layers is not zero and decays with distance from the superconductor [40]. The characteristic lengthscale for this decay is the normal-state coherence length $\xi_N = \sqrt{\hbar D/2k_B T}$, and for $s \ll \xi_N$ up to $s \approx \xi_N$ supercurrents can flow directly between the superconducting layers. However, due to the proximity effect affecting the order parameter, the transition temperature changes with the distance $s$ even if the properties of the superconducting layers remain constant [21].

The **superconducting transition temperature** $T_c$ can be analyzed to determine the inter-layer coupling. In order to analyze quantitatively the effect on NbSe₂-layer separation on $T_c$, we treat the ferecrystals as superconductor/normal metal (S/M) superlattices in the Cooper limit, where the superconducting and normal coherence lengths are larger than the layer thicknesses. Due to the charge transfer between NbSe₂ and SnSe, the SnSe layers are non-degenerate semiconductors. The superconducting transition temperature is given by [21-22]

$$T_c = 1.14\,\Theta_D \cdot \exp\left(-\frac{1}{\rho}\right) = 1.14\,\Theta_D \cdot \exp\left(-\frac{1}{N_S V_S} - \frac{N_M}{N_S}\frac{1}{N_S V_S}\frac{t_M}{t_S}\right) =$$

$$= T_{c,\text{bulk}} \exp\left(-\frac{N_M}{N_S}\frac{1}{N_S V_S}\frac{t_M}{t_S}\right), \tag{4.2}$$

where $t_M = t_{SnSe} = m \cdot 0.58125$ nm and $t_S = t_{NbSe_2} = 0.6675$ nm, and $N_{S,M}$ is the density of states (DOS) at the Fermi level and $V_{S,M}$ the pairing potential in NbSe₂ and SnSe, respectively.

This can be written as follows

$$T_c(s) = T_{c0} \cdot \exp\left(-\frac{s}{s_0}\right), \tag{4.3}$$

where

$$T_{c0} = 1.14\,\Theta_D \cdot \exp\left(-\left[1 - \frac{N_M}{N_S}\right]\frac{1}{N_S V_S}\right) \tag{4.4}$$

and

$$\frac{1}{s_0} = \left[\frac{N_M}{N_S}\right]\frac{1}{N_S V_S}\frac{1}{t_{NbSe_2}}. \tag{4.5}$$

Under the assumption that $N_M$, $N_S$ and $V_S$ do not depend on $s$, a fit to our experimental data yields (see Fig. **6**(a))

$$T_{c0} = 3.07 \text{ K}, \qquad s_0 = 2.52 \text{ nm}.$$

An extrapolation to $m \to 0$ ($s = t_{NbSe_2}$) yields for the transition temperature of a NbSe₂ monolayer

$$T_{c,ML} = 2.38 \text{ K}.$$

This is lower than the transition temperature of exfoliated and encapsulated NbSe₂ ML [18] and of macroscopic capped NbSe₂ ML [19], $T_c \approx 3.0$ K and 3.5 K, respectively.

The values of the fit parameters need to be checked for consistency with the assumptions. For instance, it is assumed that the intra-layer pairing is weak, $N_S V_S \ll 1$ [40], and that the DOS in SnSe, a non-degenerate semiconductor, is much smaller than the DOS in NbSe₂, a semimetal, with $N_M/N_S \ll 1$. From $s_0$ we obtain $\left[\frac{N_M}{N_S}\right]\frac{1}{N_S V_S} = 0.265$, and from $T_{c0} = 3.07$ K and $\Theta_D \approx 230$ K [28] we obtain

$$N_S V_S = N_{NbSe2} V_{NbSe2} \approx 0.212,$$

and

$$\frac{N_M}{N_S} = \frac{N_{SnSe}}{N_{NbSe2}} \approx 5.6 \%.$$

Therefore, both assumptions are consistent with the results. In addition to the results given by Ref. 27 for [NbSe₂]₂ bilayers, photoemission studies of $[(SnSe)_{1+\delta}]_m[NbSe_2]_1$, i. e. NbSe₂ monolayers, could provide final support.

The **cross-plane GL coherence length** $\xi_c(0)$ of the NbSe₂ layers and the **normal coherence length** $\xi_N$ of the SnSe layers are compared to the respective layer thicknesses, $t_{NbSe_2}$ and $t_{SnSe}$, in order to check whether the films are in the Cooper limit. The results are shown in Fig. **6**(b). The normal, diffusive coherence length near $T_c$ is [37]

$$\xi_N(T_c) = \sqrt{\frac{\hbar D}{2 k_B T}}\bigg|_{T=T_c}, \qquad (4.6)$$

with $m^*_{SnSe} = 0.75\, m_0$ [50]. For $m = 1$ to 3 the ratio is $\xi_N(T_c)/t_{SnSe} > 10$, corresponding to regime I; for $m = 5$ to 9 the ratio is $10 > \xi_N(T_c)/t_{SnSe} > 1$, corresponding to regime II. Therefore, the Cooper limit is satisfied for the ferecrystals in regimes I and II, with $m \leq 9$, where also a macroscopic superconducting state is established at a finite temperature. This indicates that the proximity effect in the SnSe layer is a key ingredient for the global superconducting state.

Extrapolating the trend of $\xi_N(T_c)$ with increasing $s$, it is expected that for $m > 9$ $\xi_N(T_c) < t_{SnSe}$ and the inter-layer coupling between the superconducting NbSe₂ layers is not directly mediated by proximity effect, but dominated by Josephson tunneling, a coupling that becomes exponentially weaker with increasing separation [40].

**Effective transport parameters in the normal state** are determined in order to test the assumption that the (normal-state) density of states at the Fermi level in the NbSe₂ and SnSe layers does not depend on the repeat unit thickness. If the DOS does not change, then the bandstructure of the SnSe layers does not change with thickness and the charge carrier density of the NbSe₂ and SnSe layers should be constant. In this case a one-band/one-carrier analysis of resistivity and Hall coefficient follows a simple model of parallel conductances and yields the following effective charge carrier density

$$p_{\text{eff}} = \frac{1}{(t_1+t_2)} \frac{(p_1 \mu_1 t_1 + p_2 \mu_2 t_2)^2}{p_1 \mu_1^2 t_1 + p_2 \mu_2^2 t_2} = \frac{1}{s} \frac{[p_S \mu_S t_{NbSe_2} + p_M \mu_M (s - t_{NbSe_2})]^2}{p_S \mu_S^2 t_{NbSe_2} + p_M \mu_M^2 (s - t_{NbSe_2})} \qquad (4.7)$$

and effective Hall mobility

$$\mu_{\text{eff}} = \frac{p_1 \mu_1^2 t_1 + p_2 \mu_2^2 t_2}{p_1 \mu_1 t_1 + p_2 \mu_2 t} = \frac{p_S \mu_S^2 t_{NbSe_2} + p_M \mu_M^2 (s - t_{NbSe_2})}{p_S \mu_S t_{NbSe_2} + p_M \mu_M (s - t_{NbSe_2})} \qquad (4.8)$$

This model yields a power-law dependence of the effective carrier density on the repeat unit thickness. However, the measured effective charge carrier density depends exponentially on the repeat unit thickness (see Fig. **6**(c)), which indicates that the DOS at the Fermi level changes with SnSe thickness. This is consistent with a change to the bandstructure of the SnSe layers [47, 51], with a shift of the Fermi level relative to the bandstructure and charge transfer between SnSe and NbSe$_2$ [28].

Nevertheless, the effective transport parameters can yield some information about the macroscopic behavior of the films in their normal state. For instance, the effective mean free path $l_e$

$$l_e = v_F \tau = \frac{\hbar}{e} k_F \mu_{\text{eff}} \quad (4.9)$$

and the Fermi wavevector $k_F$ (for a 3D metal)

$$k_F = (3\pi^2 \, p_{\text{eff}})^{\frac{1}{3}} \quad (4.10)$$

can be determined from the measurements, but do not contain information about the anisotropy of the transport.

The **Ioffe-Regel criterion** for a metallic-like conductor is [34]

$$k_F \cdot l_e \gg 1.$$

Its application to the $[(\text{SnSe})_{1+\delta}]_m[\text{NbSe}_2]_1$ ferecrystals is depicted in Fig. **6**(d), where it shows that for $m = 1 - 4$ (regime I) the films are "good" metals, with $k_F l_e > 3$. For $m = 5 - 9$ (regime II) the films behave like "dirty" metals, with $3 > k_F l_e > 0.3$. For $m > 9$ (regime III) they behave like "bad" insulators, with $0.3 > k_F l_e$. This is consistent with the observed temperature-dependent sheet resistance from $m = 1$ to $m = 15$, and indicates that the normal-state transport properties of these films are determined by disorder. Increasing the repeat unit thickness, the effect of disorder is increased and the films undergo in the normal-state a transition from metal to insulator at $m \approx 9$.

The **effective density of states** $N_{\text{eff}}$ can be estimated for diffusive transport with the diffusion constant $D_{\text{eff}} = \frac{1}{3} v_F l_e$ [52]

$$N_{\text{eff}} = [e^2 \, \varrho_0 \, D_{\text{eff}}]^{-1}, \quad (4.11)$$

where $\varrho_0$ is the normal-state resistivity in zero magnetic field. For the ferecrystals $N_{\text{eff}}$ decrease approximately linearly with increasing $s$ (see Fig. **6**(d)). Assuming that $m^* \approx m^*_{NbSe_2} = 0.62\, m_0$ [53], an extrapolation finds $N_{\text{eff}}(s = t_{NbSe_2}) \approx 4.3 \times 10^{27}\ \text{eV}^{-1}\text{m}^{-3} \approx 0.5 \cdot N_{2H-NbSe_2}$. The value of $N_{2H-NbSe_2} \approx 9.8 \times 10^{27}\ \text{eV}^{-1}\text{m}^{-3}$ is obtained from the electronic specific heat $\gamma \approx 38\ \text{J/m}^3\text{K}$ [20] using $\gamma = \frac{\pi^2}{3} k_B^2 \cdot N_{2H-NbSe_2}$ [54].

The **pairing potential** $V_{NbSe_2}$ and the effective DOS $N_{\text{eff}}$ determine the transition temperature [40]

$$T_c = 1.14\, \Theta_D \exp\left(-\frac{1}{N_{\text{eff}} V_{NbSe_2}}\right). \quad (4.12)$$

The pairing potential $V_{NbSe_2} = 2.84 \times 10^{-29}\ \text{eVm}^3$ for bulk 2H-NbSe₂ is obtained using the value of $N_{2H-NbSe_2}$ above and the value of $\Theta_D = 230$ K. We obtain consistently and significantly smaller values for $T_c$ than observed. Agreement between calculated and measured values can be obtained if $V_{NbSe_2}$ increases linearly with $s$, from $V_{NbSe_2} = 4.85 \times 10^{-29}\ \text{eVm}^3$ to $V_{NbSe_2} = 1.15 \times 10^{-30}\ \text{eVm}^3$. The pairing potential $V_{NbSe_2}$ is affected by screening [40], which depends the DOS at the Fermi level [54], so the pairing potential $V_{NbSe_2}$ tends to increase with decreasing $N_{\text{eff}}$.

A **cross-over from 3D toward quasi-2D superconductivity** can be inferred from our results. Our analysis highlights the role of the SnSe layers as spacer between the superconducting NbSe₂ layers, whose thickness controls the proximity-induced inter-layer coupling. Based on the transport properties in zero magnetic field, for $m > 9$ the NbSe₂ layers are decoupled and in the 2D limit, despite the fact that the 2D criterion (4.1) indicates a decoupling for $m > 18$. An explanation is that the criterion assumer weak inter-layer coupling, but for $m \leq 9$ the proximity effect in the SnSe layers enhances the inter-layer coupling and for $m > 9$ the normal-state coherence length in the SnSe layers is smaller than the spacer thickness, effectively reducing the inter-layer coupling to Josephson coupling.

However, macroscopic 2D superconductivity competes with disorder and for typical superconductor-insulator transitions in granular or disordered films [55] a normal-state sheet resistance close to $h/4e^2 \approx 6.4$ kΩ is a critical value for the transition. This value is nearly found in the sheet resistance per NbSe$_2$ layer for the $m = 6$ ferecrystal (see Fig. **2**). However, fluctuations of the resistance in parallel magnetic fields for $m = 9$ indicate a percolating path, partly in the NbSe$_2$ layers and partly through the SnSe layer. The non-linearity of the $IV$-curves and the quasi-reentrant behavior of $R_s(T)$ are strong signatures of the presence of superconducting regions in the NbSe$_2$ layers.

**Origins of the disorder** in these systems are twofold. First, the ferecrystals are textured polycrystals with grain size of the order of 50 nm, much larger than the effective mean free path $l_e$. So, if the grains in the NbSe$_2$ monolayers had a weak inter-grain coupling, it would result in a thermally activated transport: $R_s \propto \exp(E_{\text{act}}/k_B T)$ [56]. However, the sheet resistance for $m \geq 12$ is dominated by variable-range hopping: $R_s \propto \exp(\sqrt{T_{VRH}/T})$, with tunneling between localized states. This excludes grains as origin for localization. Second, because of the turbostratic disorder, each region of a NbSe$_2$ monolayer has different orientation to nearby NbSe$_2$ regions and to the adjacent SnSe layers. The SnSe layers themselves are slightly anisotropic in the $ab$ plane [28], therefore for each NbSe$_2$ monolayer region the adjacent SnSe layers present different bandstructures, which results in changes to the charge transfer from region to region. Over macroscopic lengthscales the NbSe$_2$ monolayers are homogeneous, but there are local variations in the potential landscape resulting in a spatial variation of the charge carrier density. The coupling between these regions is weak enough for transport to occur via hopping.

# Conclusions

A cross-over from metal/superconductor to insulator/superconductor is demonstrated for artificial multilayers of $[(\text{SnSe})_{1+\delta})]_m[\text{NbSe}_2]_1$ with $m$ equal 1 to 15. Our analysis highlights the role of the SnSe spacer layers: for $m = 1 - 4$ the inter-layer coupling is enhanced by proximity effect in the SnSe layer; for $m = 5 - 9$ the proximity-enhanced inter-layer coupling is reduced due to disorder in the spacer layer; and for $m > 9$ the inter-layer coupling is due to Josephson tunneling. The transport experiments lead to the conclusion that increasing the thickness of the SnSe layers changes the charge transfer to the NbSe$_2$ layers, through a change in the SnSe bandstructure and a corresponding change of the density of states in the spacer layer. Based on the transport properties, for m > 9 the NbSe$_2$ layers are decoupled and in the quasi-two-dimensional limit. The observed charge transfer between constituents, combined with the ability to change the identity, thickness and alloy of the rock salt constituent, provides the general opportunity to control the superconductive inter-layer coupling over a wide material basis.

# Acknowledgments

OC, KM, TUG, CG and SFF thank Martina Trahms for measurements and scientific discussions, and Linus Grote for scientific discussions. OC, KM, TUG, CG and SFF acknowledge support from the Deutsche Forschungsgemeinschaft under grant Inst 276/709-1. MBA, KH and DCJ acknowledge support from the National Science Foundation under grant DMR-1710214.